\pdfoutput=1

\documentclass[]{raa}

\usepackage{mathptmx}
\usepackage{ae,aecompl}

\usepackage{graphicx,times}	
\usepackage{amsmath}	
\usepackage{amssymb}	
\usepackage{pifont}
\usepackage{txfonts}
\usepackage{url}
\usepackage{subfigure}
\usepackage{longtable}
\usepackage{lscape}
\usepackage{multirow}
\usepackage{color}
\usepackage{textcomp}
\usepackage{natbib}
\usepackage{epstopdf}
\usepackage{pdfpages}
\usepackage[colorlinks,linkcolor=red,anchorcolor=green,citecolor=blue]{hyperref}
\usepackage{soul}
\usepackage{appendix}

\begin{document}

\title{ Ultra-wide Bandwidth Observations of 19 pulsars with Parkes telescope}


   \volnopage{Vol.0 (2021) No.0, 000--000}      
   \setcounter{page}{1}          
    \author{Zu-Rong Zhou
      \inst{1,2}
   \and Jing-Bo Wang
      \inst{1,3,4}
   \and Na Wang
      \inst{1,3,4}
      \and George Hobbs
      \inst{5}
   \and Shuang-Qiang Wang
      \inst{1,2,6}
   }

 \institute{Xinjiang Astronomical Observatory, Chinese Academy of Sciences, Urumqi, XinJiang 830011, China \\
        \and
            University of Chinese Academy of Sciences, 19A Yuquan Road, Beijing 100049, China \\
       \and
            Xinjiang Key Laboratory of Radio Astrophysics, 150 Science1-Street, Urumqi, Xinjiang, 830011, China\\
       \and
            Key Lab of Radio Astronomy, Chinese Academy of Sciences, Beijing 100101, China\\
       \and
            CSIRO Astronomy and Space Science, PO Box 76, Epping, NSW 1710, Australia\\
       \and
           CAS Key Laboratory of FAST, NAOC, Chinese Academy of Sciences, Beijing 100101, People's Republic of China\\
\vs\no
   {\small Received 20xx month day; accepted 2022 May 10th}}
\email{{\it wangjingbo@xao.ac.cn; na.wang@xao.ac.cn}}

\abstract{Flux densities are basic observation parameters to describe pulsars. In the most updated pulsar catalog, 24$\%$ of the listed radio pulsars have no flux density measurement at any frequency. Here, we report the first flux density measurements, spectral indices, pulse profiles, and correlations of the spectral index with pulsar parameters for 19 pulsars employing the Ultra-Wideband Low (UWL) receiver system installed on the Parkes radio telescope. The results for spectral indices of $17$ pulsars are in the range between $-0.6$ and $-3.10$. The polarization profiles of thirteen pulsars are shown. There is a moderate correlation between the spectral index and spin frequency. For most pulsars detected, the S/N ratio of pulse profile is not high, so DM, Faraday rotation measure (RM), and polarization can not be determined precisely. Twenty-nine pulsars were not detected in our observations. We discuss the possible explanations for why these pulsars were not detected.
}

   \authorrunning{Z. R. Zhou et al. 2021}                     
   \titlerunning{Wide Bandwidth Observations of 19 pulsars}       

   \maketitle


\section{Introduction}\label{Sect.1}

Pulsars have been discovered for more than 53 years \citep{Hewish1968}, its radiation mechanism has not yet been fully understood ~\citep{Jankowski2017}. One of the widely accepted models is that the radio emission regions are limited to the open polar cap inside the light-cylinder radius ~\citep{Ruderman1975}. However, the specific radiation area, physical process, and other details about pulsar radio emission are still unclear. Studying the radio spectra of pulsars is helpful to understand their emission mechanism. Unfortunately, only a few hundred pulsars have been determined spectra. Only a tiny proportion of pulsars have been studied in a relatively wide frequency range (e.g.,~\citealt{Dai2015}).

Measurement of flux density at multiple frequencies are needed when we determine the spectra of pulsars. The Australia Telescope National Facility (ATNF) pulsar catalog ~\citep{Manchester2005} provides a database for pulsar observational parameters. Pulsar flux densities in this catalog are well-known near 1400 and 400 MHz relatively, where most of the pulsars were found, but few are known at other frequencies. We have used the most updated catalog, version 1.64, and out of the 2872 radio pulsars in the catalog, 680 do not have flux density values in any radio waveband. And, 97.4$\%$ have no historical flux density measurements near 2 GHz, 74.3$\%$ have no recorded flux density values near 400 MHz, 32.5$\%$ have no recorded measurements of flux density close to 1400 MHz. The pulsar parameters in the catalog are collected by different telescopes and different generations of receivers and backends, and each has its system error. Significant differences were found between multiple measurements ~\citep{Levin2013}. Therefore, it is vital to obtain the absolute flux density calibrated measurements of pulsars ~\citep{Jankowski2017}. Efforts to measure radio spectra began in earnest with ~\citet{Rankin1970} and ~\citet{Sieber1973}, followed by ~\citet{Malofeev1980} and ~\citet{Izvekova1981}. They measured the spectra at low frequencies near 100 MHz and below, finding that most pulsars have cliffy spectra and can be expressed as a simple power-law. Some pulsar spectra deviate from a simple power-law at low frequencies and show a turn-over~\citep{Rankin1970}, while some show a high-frequency cut-off in the form of a spectrum steepening or a break in the spectrum ~\citep{Sieber1973}. ~\citet{Lorimer1995} has studied the spectra of 280 pulsars and obtained a mean spectral index of $-1.6$. A study of the spectral properties of 441 radio pulsars observed with the Parkes telescope ~\citep{Jankowski2017} and found about $79\%$ of the pulsars could be classified as simple power-law spectra.

People realized that pulsars were usually highly polarized long ago ~\citep{Lyne1968}. The long-term stability of the pulse profile and its complexity in total intensity and polarization are also important characteristics of pulsars. The mean pulse profile and polarization properties help understand the geometry of the star, the pulse emission mechanism, and the beaming of pulsar radiation. Mean pulse profiles often have double or triple components, leading to different description origin of the beaming of the emission ~\citep{Backer1976}, and the spectral index often varies from one component to the next (e.g.,~\citealt{Backer1972}). Polarization properties of pulsars are normally described in terms of the four Stokes parameters. Many pulsars show a systematic variation across the pulse profile of the position angle (PA). The observed PA swings in many pulsars like the "S"-shaped curve, explained by the rotating vector model (RVM, ~\citealt{Radhakrishnan1969}). Of course, the observed PA variations are not always continuous and smooth. Both normal pulsars and millisecond pulsars (MSPs) are often can be seen in a high degree of linear polarization, and orthogonal-mode PA jumps.(see e.g. ~\citealt{Yan2011}). Linear polarization is relatively stronger than circular polarization usually. Circular polarization is most often with a sense reversal near the profile mid-point, usually associated with the core or central component of the profile ~\citep{Rankin1983}.

In addition to the shape properties of the mean pulse profile, the width of the mean pulse profile is also used to manifest the profile characteristics. The angular width of the mean pulse is also known as the width of the observed beam. $W_{50}$ and $W_{10}$ are pulse widths that are often mentioned and studied (see e.g., ~\citealt{Gould1998}). Pulsars exhibit a diverse frequency dependence of average pulse profile. ~\citet{Phillips1992} shows the whole profile width and spacing between each component of three pulsars increase with the decrease of observation frequency from 4800 MHz to 50 MHz using the Arecibo 305-m radio telescope. The increasing pulse width trend is very obvious in the frequency band below 1GHz. ~\citet{Xilouris1996} extended the measurements of profile width to 32 GHz and combined their measurements with published values at lower frequencies, showing a better fit with $W_{50}$ of six pulsars. One hundred fifty normal pulsars, the $W_{10}$ of which are well fitted versus the observing frequencies with the ~\citeauthor{Thorsett1991} relationship ~\citep{Chen2014}.

 Lately, Parkes 64m radio telescope has been equipped with an ultra-wideband, low-frequency receiver. The frequency range of the receiver system is from 704 to 4032 MHz continuously. Besides, the receiver has excellent sensitivity and polarization properties ~\citep{Zhang2019}. The high-sensitivity, well-calibrated and wide frequency observations with Parkes telescope could efficiently provide us a systematic and uniform sample of pulsar flux densities ~\citep{Hobbs2020}. In this work, we describe observations of 19 pulsars using the UWL receiver system in Section 2. Our results are shown in Section 3. We discuss the results and give a concluding summary in Section 4.

\section{Observations and Data Reduction}\label{Sect.2}

We have used the ATNF pulsar catalog version 1.59 to identify all radio pulsars without a flux density measurement near 1400 MHz that can be observed by the Parkes telescope. This list provides 505 pulsars. Then we remove all pulsars only detected in a survey that has a poorly defined sensitivity limit (for instance, some surveys are simply defined in the catalog as "miscellaneous" and very long observations with large telescopes may have detected the pulsars). After applying this selection, we have 327 pulsars remaining. Based on the known survey sensitivities, we know we can obtain a detectable signal close to 1400 MHz with observation durations of 30 minutes for all these pulsars. Still, some will be detectable at a high signal-to-noise ratio (S/N) with significantly shorter integration times. In particular, 54 of these pulsars were discovered in the Molonglo surveys or with the Parkes 70 cm receiver ~\citep{Manchester1978}. Such pulsars can be detected with only a few minute integration time.

These 54 pulsars were observed with Parkes using the UWL receiver. All the details of the receiver and backends are described in ~\citet{Manchester2013} and ~\citet{Hobbs2020}. Medusa and Parkes Digital Filter Banks 4 (PDFB4) are simultaneously used for the signal pre-processor systems in the observation. These frequencies of two backends are centered at 2368 MHz and 1369 MHz, respectively. The total recording bandwidth of Medusa (3328 MHz) was subdivided into 3328 frequency channels, while PDFB4 of 256 MHz was subdivided into 1024 frequency channels. In the meantime, each pulsar period was divided into 1024 phase bins with each band. The data for all observations were recorded in a ¡°fold¡± mode in 30 seconds and de-dispersed online coherently. Note that PDFB4 has no flux calibrator.

The data were processed with {\sc psrchive} software package ~\citep{Hotan2014}. We removed $5\%$ of the band edges and manually excised data affected by narrow-band and impulsed radio frequency interference (RFI). We used the {\sc psrchive} program {\sc paz} to eliminate RFI automatically. Then we used {\sc pazi} to inspect the pulse profiles visually and remove sub-integrations or frequency channels affected by RFI manually. We also used observations of the radio galaxy 3C 218 (Hydra A) to transform the measured intensities to absolute flux densities, using on- and off-source pointing to measure the obvious brightness of the noise diode as a function of radiofrequency. The pulsar observations were calibrated using their associated calibration files using {\sc pac} to transform the polarization products to Stokes parameters, to flatten the band-pass, and to calibrate the pulse profiles in flux density units. We formed a noise-free standard template from our observations using {\sc paas} and then using {\sc psrflux} to obtain the average flux density. All raw data of 54 pulsars can be downloaded from CSIRO data archive~\citep{Hobbs2011}.

\begin{table*}
\begin{center}
\caption[]{Pulsar Observational Parameters\label{t1}}
\setlength{\tabcolsep}{2.5pt}
\small
 \begin{tabular}{cccccccccc}
  \hline\noalign{\smallskip}
 PSR J & Period & DM & Frequency & MJD & Nobs &Length & S/N & $S_{2368}$ & $W_{50}$\\
 $\qquad$ & (s) & (cm$^{-3}$pc)&(MHz)&& & (min)& & (mJy) & (deg)\\
\hline\noalign{\smallskip}
J0057$-$7201&0.74&27 &1369&58744&1&9.989&7.56& & \\
J0348$+$0432 & 0.04 & 40.46&2368 &58744&1& 10.253 & 16.465 & 0.43(2)&4(1) \\
J0418$-$4154 &0.76 & 24.54&1369/2368&58744& 1& 10.268 & 22.239 & 0.39(2)&6.2(7) \\
J0458$-$0505&1.88&47.81 &1369&58744&1&9.977 &8.34& & \\
J1057$-$4754 &0.63& 60.00 &1369/2368 &58739&1& 10.253 & 19.216 & 0.38(17)&7.4(7) \\
J1157$-$5112&0.04 &39.67 &1369&58767 &4 &101.998&51.59& & \\
J1420$-$5416 & 0.94 &129.60&2368&58740&1 & 10.284 & 35.131 & 1.15(4)&8.3(9) \\
J1423$-$6953 & 0.33 & 123.98&1369/2368&58739&1 &10.319 & 17.070 & 0.29(2)&3.7(9) \\
J1510$-$4422&0.94&84&1369&58740&1 &9.974&12.33& & \\
J1527$-$3931 & 2.42 & 49.00&1369/2368&58740 &1& 10.385 & 32.450 & 0.95(5)&4(2) \\
J1539$-$6322&1.63 &163.5 &1369&58740&1 &9.949&21.39& & \\
J1604$-$7203&0.34&54.37&1369&58740&1 &9.992 &18.48& & \\
J1615$-$2940 & 2.48 & 44.79&1369/2368&58740& 1& 10.369 & 19.795 & 0.83(4)&3(2)\\
J1625$-$4048 & 2.36 &145.00&1369/2368&58740&2 & 20.771 & 54.222 & 1.08(4)&5(1)\\
J1721$-$1939 & 0.40 & 103.00&2368&58740& 1& 10.252 & 15.754 & 0.23(17)&9.1(9)\\
J1728$-$0007 &0.39 & 41.09&1369/2368&58740&1 & 10.402 & 31.494 & 0.54(17)&17.2(7)\\
J1746$-$2856 &0.95 & 1168.00&2368&58740&1& 10.368 & 16.757 & 0.80(3)&21(1) \\
J1805$+$0306 &0.22 & 80.86 &1369/2368&58740&1& 10.253 & 37.390 & 0.78(2)&11.7(6)\\
J1810$-$5338 &0.26 & 45.00&1369/2368&58742&2&20.538 & 267.851 & 2.05(17)&15.0(5) \\
J1816$-$5643 & 0.22 & 52.40&2368&58740&1& 10.418 & 14.195 & 0.30(17)&8.2(9) \\
J1833$-$6023 & 1.89 & 35.00&1369/2368&58744&1& 10.418 & 65.036 & 1.47(5)&11(1)\\
J1848$-$1952 & 1.06 & 18.23&1369/2368&58740&1& 13.059 & 44.213 & 0.95(7)&5(2) \\
J1854$-$1421 &1.15 & 130.40&1369/2368&58740&1 & 10.385 & 75.549 & 2.06(6)&9(1) \\
J1903$-$0632 &0.43 & 195.61&1369/2368&58740&1& 10.400 & 52.803 & 1.24(4)&7.0(6)\\
J2222$-$0137 &0.03 & 3.28&1369/2368&58744&1 & 10.253 & 124.332 & 1.41(5)&6.6(6)\\
\noalign{\smallskip}\hline
\end{tabular}
\end{center}
\end{table*}

\begin{table*}
\begin{center}
\caption[]{Flux Densities and Spetral Indices for 18 Pulsars\label{t2}}
\setlength{\tabcolsep}{0.6pt}
\small
 \begin{tabular}{ccccccccccccccl}
  \hline\noalign{\smallskip}
PSR J &$S_{150}$&$S_{200}$&$S_{300}$& $S_{400}$ & $S_{600}$&$S_{700}$ & $S_{800}$ & $S_{1400}$ & $S_{2000}$ & $S_{3000}$ &$S_{6000}$ & b & $\alpha$ & References\\
$\qquad$&(mJy) &(mJy) &(mJy) &(mJy) &(mJy)&(mJy)&(mJy)& (mJy) & (mJy)&(mJy)&(mJy)& & &for position\\
  \hline\noalign{\smallskip}
J0348$+$0432&10.8 & && & &  & 1.8 &0.34(2)& & & &$1(1)$&$-1.21(43)$&\citet{Lynch2013}\\
J0418$-$4154&&40(3)&10.3&&&&2.2(7)&&&&&0.1(2)&$-3.10(4)$&\citet{Bhattacharyya2016}\\
J1057$-$4754& &&& & & &  &0.52(3)&0.46(2)&0.32(2)&&0.54(3)&$-0.6(1)$&This work\\
J1420$-$5416& &&&9& &&4.6(4)&1.61(5)&0.46(3)& & &$1.92(25)$&$-1.27(19)$&\citet{Taylor1993}\\
J1423$-$6953& &&& & & &  &0.32(3)&0.28(2)& & &0.32& $-0.37$& This work\\
J1527$-$3931& &&& 11& & &2.8(4) &0.92(5)& & & &1(8)&$-2(3)$& \citet{Taylor1993}\\
J1615$-$2940& &&& 3.1(5)& 2.2& & &0.39(4)& & & &0.4(4)&$-1(1)$&\citet{Lorimer1995}\\
J1625$-$4048& &&&17& &3.3(8)& &1.01(3)& & & &0.5(8)&$-2.74(27)$&\citet{Manchester1996}\\
J1728$-$0007& &&&11(1)&4.1& & &0.65(2)&0.25(18)& & &0.6(1)&$-2.3(3)$&\citet{Lorimer1995,Lommen2000}\\
J1746$-$2856&&& & & & & &  &1.31(8)&0.76&0.067&3(9)&$-2(8)$&\citet{Johnston2006}\\
J1805$+$0306&18.7&&&5& & & &1.19(3)&1.00(3)&0.52(18)& &$1.04(80)$&$-1.29(30)$& \citet{Stokes1985}\\
J1810$-$5338&&& &12&18&12(3)& &4.08(5)&1.37(3)&0.87(17)& &42&$\alpha_{1}=1,\alpha_{2}=-2.0(3)$& \citet{Taylor1993}\\
J1816$-$5643&&& & & & & &0.30(19)&0.21(3)& & &0.3&$-0.99$&This work\\
J1833$-$6023&&& & 5.5& & & &1.49(5)& & & &1.49&$-1.04$&\citet{Taylor1993}\\
J1848$-$1952&&& & 17(2)&7& & &1.14(7)&0.96(6)& & &$1.25(82)$&$-2.08(34)$&\citet{Lorimer1995}\\
J1854$-$1421&&& & 8(1)&6.0(5)&9(2)&4(3)&3.01(7)&1.3(10)&0.91(4)& &3.85(4)&$-0.74(6)$&\citet{Lorimer1995}\\
J1903$-$0632&&& &23(1)&11.3(7)&7(2)&5.5(7) &1.99(6)&0.72(6)&0.26(2)& & 1.97(15)&$-1.98(7)$&\citet{Lorimer1995,Malofeev2000,Maron2000,Jankowski2017}\\
J2222$-$0137&11.5& && & &1(1)& 2.6(5)&2.64(7)&0.45(4)&0.46(2)& &1.17(17)&$-1.00(9)$&\citet{Boyles2013}\\
  \noalign{\smallskip}\hline
\end{tabular}
\end{center}
\end{table*}

\begin{table*}
\begin{center}
\caption[]{Correlation of spectral index $\alpha$ with $\lg|x|$ for different pulsar parameters $x$. $r_{s}$, and $p$ are the correlation coefficient of Spearman rank, the corresponding $p-$value and the number of the sample size, respectively. The correlations with a p-value of less than $5$ per cent and an absolute value of $r_{s}$ of at least $0.4$ are marked in bold. The sample size for each data set is listed above the references. NP: Normal pulsars; IP: Isolated pulsars.\label{t3}}
\setlength{\tabcolsep}{1pt}
\small
 \begin{tabular}{ccccccc}
 \hline\noalign{\smallskip}
Set &NP&IP&NP&NP&NP&NP\\
Npsr&17&15&21&267&572&323\\
$\sharp$References&This work&This work&\citet{Zhao2019}&\citet{Jankowski2017}&\citet{Han2016}&\citet{Lorimer1995}\\
\hline
$x$&$r_{s}(p)$&$r_{s}(p)$&$r_{s}(p)$&$r_{s}(p)$&$r_{s}$&$r_{s}(p)$\\
$\nu$&\textbf{0.41}(9.23e-03)&0.33(3.09e-02)&&0.37(5.7e-10)&&\\
$\dot{\nu}$&0.17(1.44e-02)&0.11(7.20e-03)&&\textbf{0.43}(3.1e-13)&&\\
$B_{LC}$&0.26(6.19e-02)&0.16(1.93e-01)&&\textbf{0.43}(2.5e-13)&&\\
$\tau_{c}$&0.33(5.53e-02)&0.23(4.78e-01)&\textbf{$-$0.57}(0.002)&$-0.39$(5.9e-11)&$-0.20$&$-0.19$(8e-4)\\
$B_{surf}$&$-0.47$(1.38e-01)&\textbf{$-$0.43}(1.45e-02)&0.02(0.95)&&0.03&\\
$\dot{E}$&0.047(2.33e-01)&0.007(7.98e-02)&\textbf{0.723}(0.001)&\textbf{0.43}(1.4e-13)&0.26&\\
$P$&\textbf{$-$0.41}(2.12e-02)&-0.33(2.47e-03)&\textbf{$-$0.718}(0.001)&&$-0.21$&$-0.22$(6e-5)\\
$\dot{P}$&-0.40(3.21e-01)&$-0.33$(1.93e-01)&0.39(0.12)&0.28(4.9e-06)&0.13&\\
  \noalign{\smallskip}\hline
\end{tabular}
\end{center}
\end{table*}

\begin{figure}
\centering
\includegraphics[width=4.1cm,angle=0]{1.jpg}
\includegraphics[width=4.1cm,angle=0]{2.jpg}
\includegraphics[width=4.1cm,angle=0]{3.jpg}
\includegraphics[width=4.1cm,angle=0]{4.jpg}
\includegraphics[width=4.1cm,angle=0]{5.jpg}
\includegraphics[width=4.1cm,angle=0]{6.jpg}
\includegraphics[width=4.1cm,angle=0]{7.jpg}
\includegraphics[width=4.1cm,angle=0]{8.jpg}
\includegraphics[width=4.1cm,angle=0]{9.jpg}
\includegraphics[width=4.1cm,angle=0]{10.jpg}
\includegraphics[width=4.1cm,angle=0]{11.jpg}
\includegraphics[width=4.1cm,angle=0]{12.jpg}
\includegraphics[width=4.1cm,angle=0]{13.jpg}
\includegraphics[width=4.1cm,angle=0]{14.jpg}
\includegraphics[width=4.1cm,angle=0]{15.jpg}
\includegraphics[width=4.1cm,angle=0]{16.jpg}
\includegraphics[width=4.1cm,angle=0]{17.jpg}
\includegraphics[width=4.1cm,angle=0]{18.jpg}
\caption{Spectra for 18 pulsars and power-law fits the data (blue lines). The red bars give the uncertainties of the flux densities at the different frequencies.
}
\label{Fig1}
\end{figure}

\begin{figure}
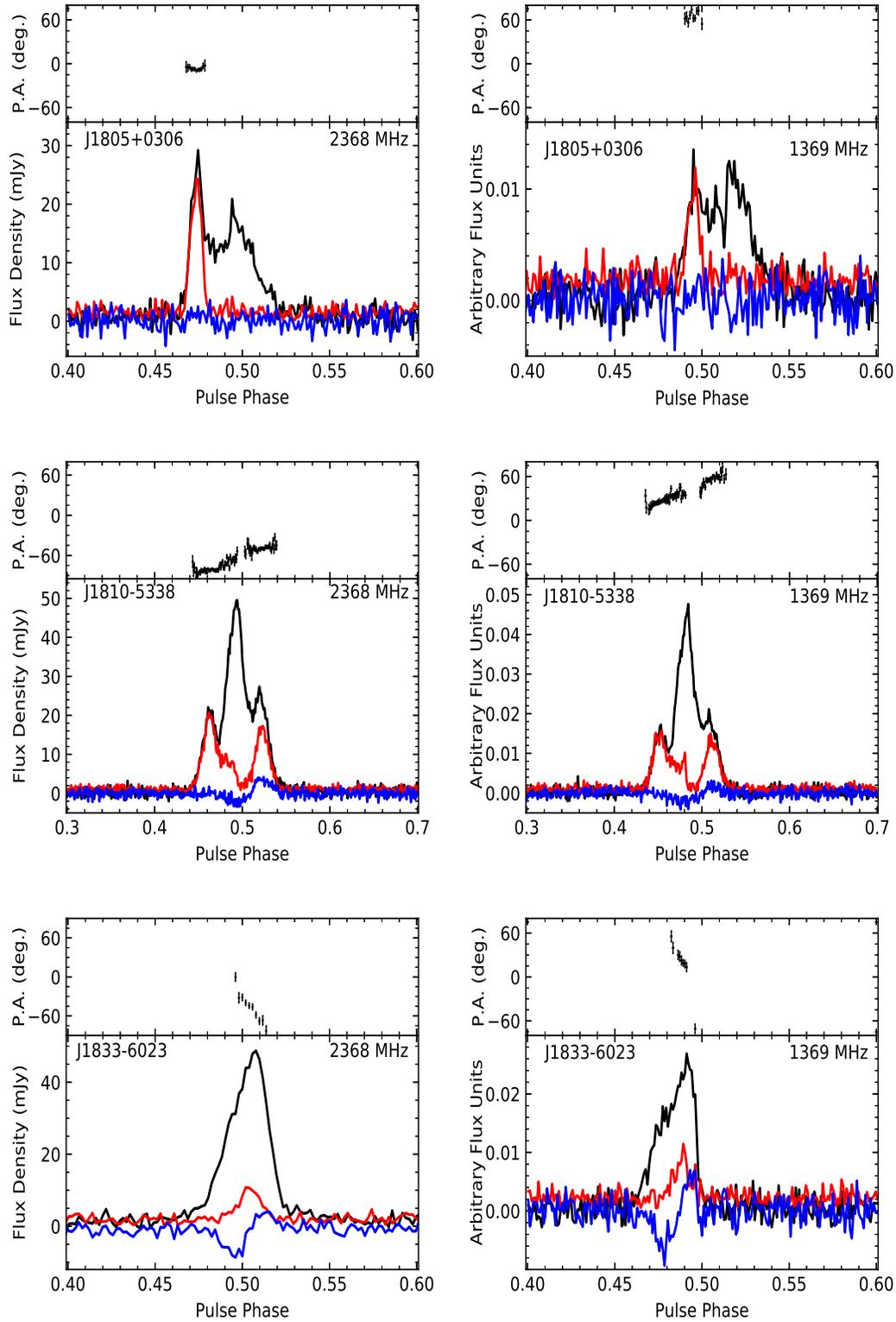

\centering
\includegraphics[width=7cm, height=7cm, angle=0]{1805m.jpg}
\includegraphics[width=7cm, height=7cm, angle=0]{1805d.jpg}
\includegraphics[width=7cm, height=7cm, angle=0]{1810m.jpg}
\includegraphics[width=7cm, height=7cm, angle=0]{1810d.jpg}
\includegraphics[width=7cm, height=7cm, angle=0]{1833m.jpg}
\includegraphics[width=7cm, height=7cm, angle=0]{1833d.jpg}
\caption{Average polarization profiles for PSR J1805$+$0306, PSR J1810$-$5338, PSR J1833$-$6023, PSR J1848$-$1952, PSR J1903$-$0632 and PSR J1854$-$1421 at 1369 and 2368 MHz. The total intensity is shown in black, linearly-polarization, and circular-polarization are shown in red and blue respectively.
}
\label{Fig2}
\end{figure}

\begin{figure}
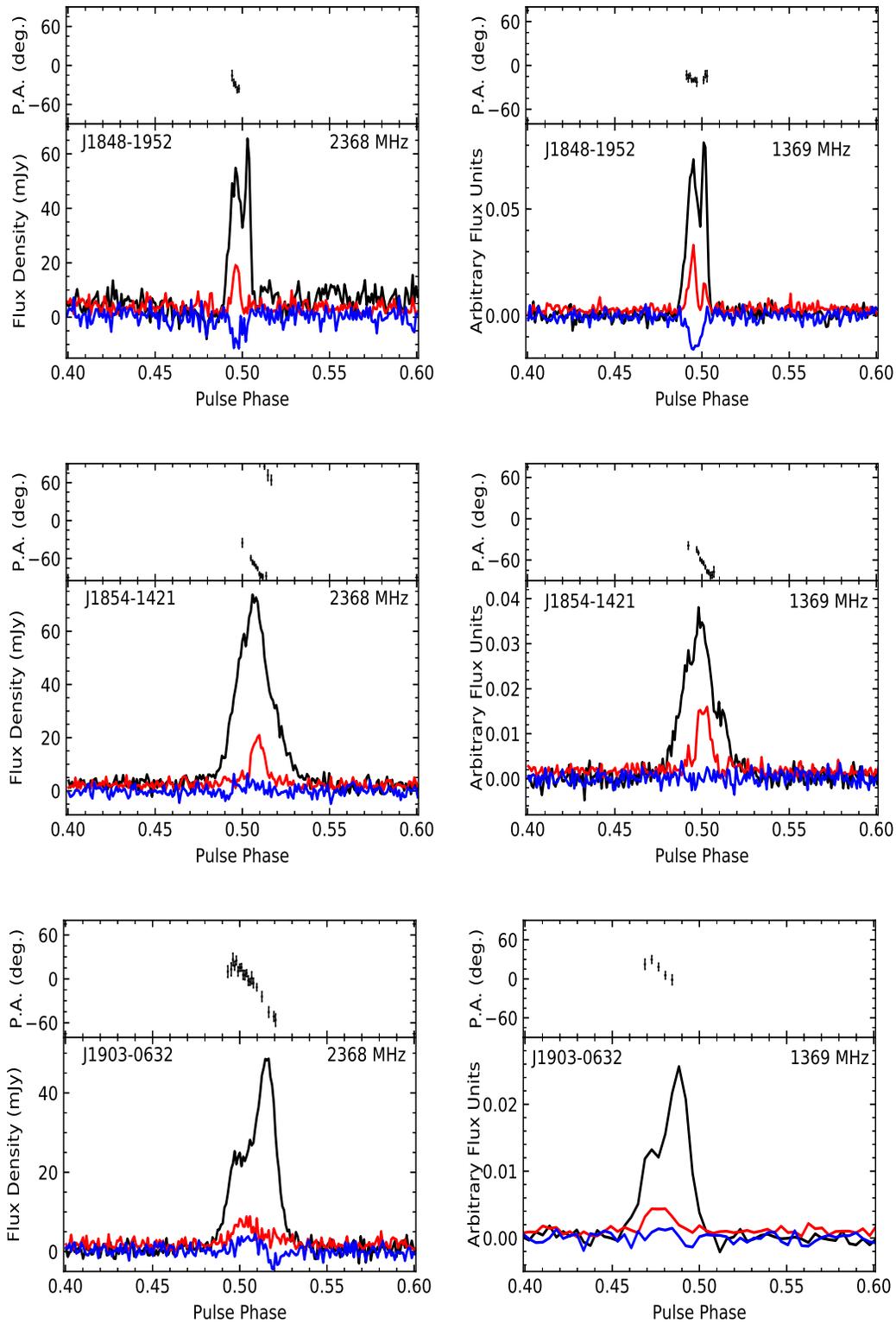

\centering
\includegraphics[width=7cm, height=7cm, angle=0]{1848m.jpg}
\includegraphics[width=7cm, height=7cm, angle=0]{1848d.jpg}
\includegraphics[width=7cm, height=7cm, angle=0]{1854m.jpg}
\includegraphics[width=7cm, height=7cm, angle=0]{1854d.jpg}
\includegraphics[width=7cm, height=7cm, angle=0]{1903m.jpg}
\includegraphics[width=7cm, height=7cm, angle=0]{1903d.jpg}
\addtocounter{figure}{-1}
\caption{-continued}
\label{Fig2}
\end{figure}

\begin{figure}
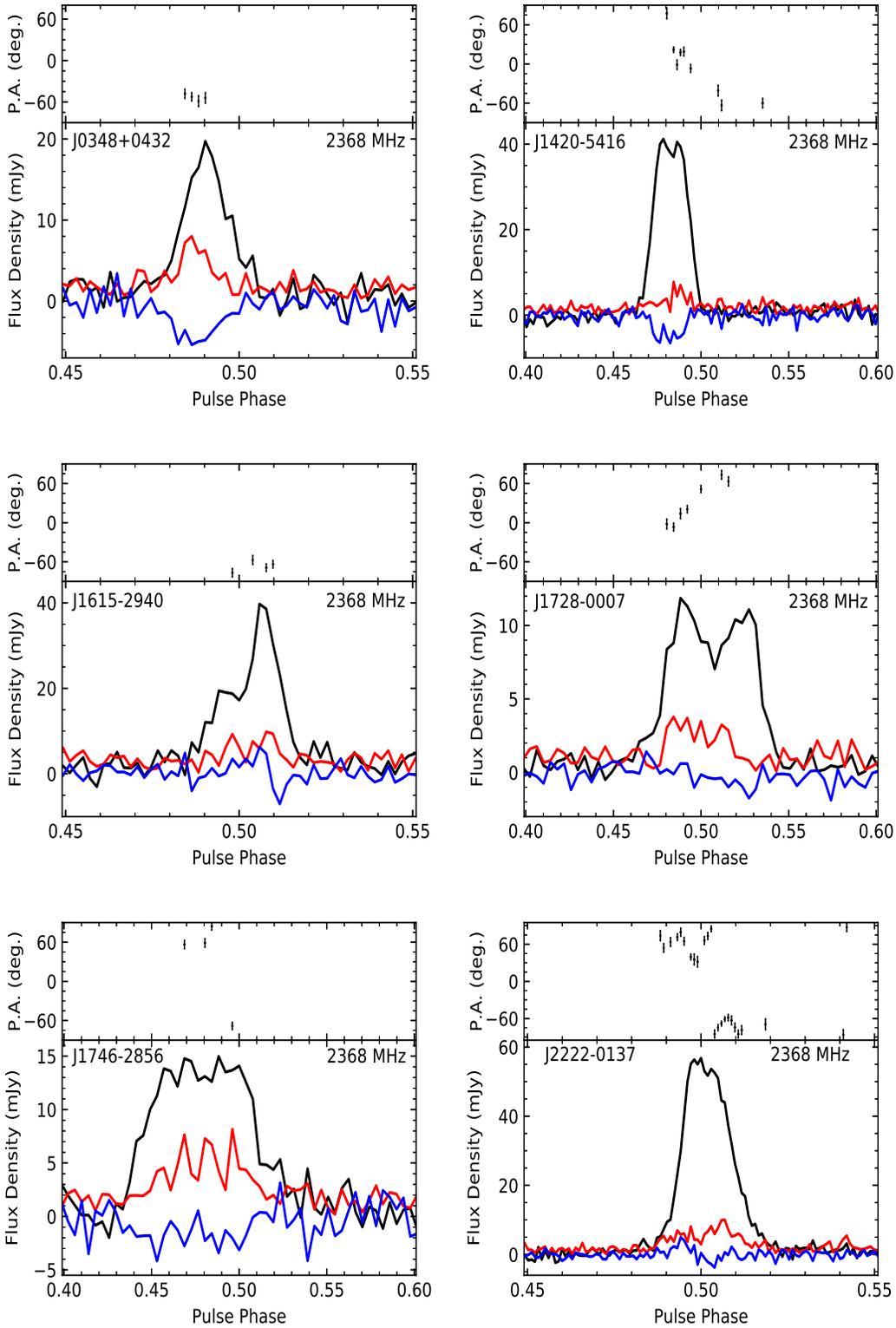

\centering
\includegraphics[width=7cm, height=7cm, angle=0]{0348m.jpg}
\includegraphics[width=7cm, height=7cm, angle=0]{1420m.jpg}
\includegraphics[width=7cm, height=7cm, angle=0]{1615m.jpg}
\includegraphics[width=7cm, height=7cm, angle=0]{1728m.jpg}
\includegraphics[width=7cm, height=7cm, angle=0]{1746m.jpg}
\includegraphics[width=7cm, height=7cm, angle=0]{2222m.jpg}
\caption{Average polarization profiles for PSR J0348$+$0432, PSR J1420$-$5416, PSR J1615$-$2940, PSR J1728$-$0007, PSR J1746$-$2856, and PSR J2222$-$0137 at 2368 MHz. The total intensity is shown in black, linearly-polarization, and circular-polarization are shown in red and blue respectively.}
\label{Fig3}
\end{figure}

\begin{figure}
\centering
\includegraphics[width=7cm, height=7cm, angle=0]{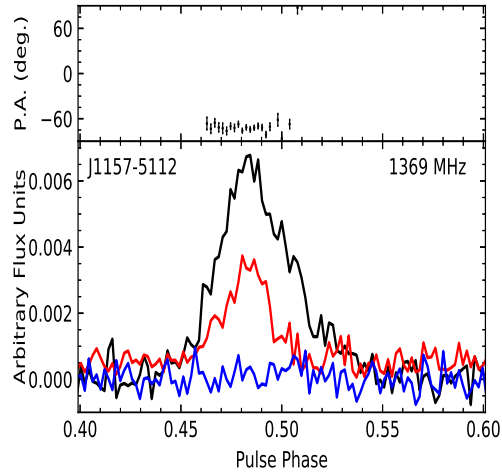}
\caption{Average polarization profile for PSR J1157$-$5112 at 1369 MHz.}
\label{Fig4}
\end{figure}

\section{Results}\label{Sect.3}

We obtained data sets of 54 pulsars for which no flux densities were previously published near 1400 MHz and have successfully detected 25 pulsars. Only 14 of 25 pulsars were detected at 1369 and 2368 MHz at the same time. We gained pulse profiles and flux densities of 19 pulsars at 2368 MHz. No flux calibrator is available for the other six pulsars at 1369 MHz.

\subsection{Flux Density Measurements}\label{Sect.3.1}

 One of the fundamental properties of any astronomical source is its flux density, so we recommend that these flux densities should be included in the next version of the pulsar catalog. Our main results are listed in Table 1, in which the pulsar name, pulse period, dispersion measure (DM), observation frequency, MJD, observation times, observation length, S/N of the pulse profile, the average flux density obtained at 2368 MHz, and the width of pulse profiles at 50 percents ($W_{50}$) are given in column order. Nineteen pulsars which are observed at 2368 MHz in Table 1 are mainly young pulsars with characteristic ages mostly between $10^{6}$ and $10^{7}$ years, except for PSR J0348+0432 and PSR J2222-0137, which are in binary systems with a white dwarf (WD) companion of different masses, respectively. The observation time of most pulsars is about 10 minutes. (The $S_{2368}$ represents flux density at 2368 MHz.) In Table 1, flux densities range from 0.23 to 2.06 mJy. Noted that $W_{50}$ ranges from $3^{\circ}$ to $21^{\circ}$. The error of $W_{50}$ was estimated by determining how the width changes when the 50 percent flux density cuts across the profile moves up or down by the baseline root-mean-square (RMS) noise level ~\citep{Dai2015}.
  $W_{50}$ and $W_{10}$ of 17 pulsars from literature at other frequencies are given in Table B.2. The $W_{50}$ for PSR J1833-6023 at two frequencies are identical. The $W_{50}$ of 10 pulsars decreases with frequency, and six pulsars increase with frequency.

\subsection{Spectral indices}\label{Sect.3.2}

 The obtained flux densities can be used to measure the spectral indices. Many pulsars have been observed at one or other frequencies of 400, 600, 800, 1400, 2000, and 3000 MHz (e.g.,~\citealp{Dai2015, Jankowski2017}). To estimate spectral indices, we divide the UWL data sets into three sub-bands and measure the flux density for each of them. Their center frequencies are close to 1400, 2000, and 3000 MHz, with sub-band widths of 400 MHz, 400 MHz, and 600 MHz, respectively. The results are listed in Table 2 as $S_{1400}$, $S_{2000}$, and $S_{3000}$. Except for six pulsars(PSR J1057$-$4754, PSR J1805$+$0306, PSR J1810$-$5338, PSR J1854$-$1421, PSR J1903$-$0632, and PSR J2222$-$0137), the flux density of other 13 pulsars cannot be measured in all the three subbands since too many frequency channels are removed. The pulse profiles of only one or two sub-bands can be seen for some pulsars. Previously published $S_{150}$, $S_{200}$,$S_{300}$, $S_{400}$, $S_{600}$, $S_{700}$, $S_{800}$, and $S_{6000}$ for these pulsars are used to estimate the spectral index, and the references are given in the last column of Table.1. Assuming a simple power law of the form $S_{\nu}= bx^{\alpha}$, where $ x = \frac{\nu}{\nu_{0}}$, $\nu$ is the center frequency and ${\nu_{0}} = 1.4$ GHz a constant reference frequency.
The fit parameters are the spectral index ${\alpha}$ and a constant $b$, and their values are provided in Table 2. The spectral behaviour of the 17 pulsars can be well described by a simple power$-$law over the frequency range considered. All of the 17 pulsars with spectral indices are between $-0.6$ and $-3.10$. For PSR J1810-5338, a broken power law is a batter to describe the spectrum with 600 MHz of cut-off frequency, $\alpha_{1}$ is the spectral index before and $\alpha_{2}$ the one after the break. The spectral plots of 18 pulsars and power-law fits the data are shown in Fig.1. According to previous statistics, the mean spectra index of pulsars with a simple power-law spectrum is about $-1.60$. The majority of our spectral indices are flatter than the mean values from ~\citet{Jankowski2017}, but our measurements are not particularly unusual as there are many other pulsars that also have similar spectral indices. As our sample is relatively small, we do not think our pulsar sample is particularly inclined to flat spectrum pulsars.

\subsection{Polarization profiles}\label{Sect.3.3}

Pulse profiles at 2368 MHz for all the 19 pulsars are presented for the first time. The profiles of 20 among 25 pulsars were published on European Pulsar Network (EPN) database near 1.4 GHz for the first time except for five pulsars (PSR J1057$-$4754, PSR J1604$-$7203, PSR J1625$-$4048, PSR J1728$-$0007, and PSR J1805$+$0306) ~\footnote{http://www.epta.eu.org/epndb}. The pulse profiles at 1369 MHz and 2368 MHz of most pulsars are similar in our observations. Most of the average pulse profiles in Fig.A1 and Fig.A2 have a single peak, and the pulse profiles are narrow. Fig.s 2$-$4 show the calibrated polarization profiles for 13 pulsars in our sample. An RM fitting during the calibration is considered during the calibration of polarization. Only six pulsars show strong polarization characteristics at two frequencies simultaneously, while the other seven pulsars only have polarization profiles at 1369 or 2368 MHz. Polarization can not be measured precisely for other pulsars due to the limited S/N ratio. Results for individual pulsars are described in detail as follows.

\subsubsection{\textbf{PSR J0348$+$0432}}
PSR J0348$+$0432, a binary pulsar in a 2.46-hour orbit with a low-mass ELL1 WD companion, was first discovered in the Green Bank Telescope (GBT) 350 MHz drift-scan survey. The pulse profile of the pulsar at 350 MHz has three components \citep{Lynch2013}, while Fig.3 shows a single peak at 2368 MHz and is linearly polarized. There is significant right-hand circular polarization and slight shallow rotation in the position angle (PA) variation across the profile.

\subsubsection{\textbf{PSR J1157$-$5112}}
PSR J1157$-$5112 was discovered in the first high-frequency survey of intermediate Galactic latitudes. It is a 44 ms pulsar and the first recycled pulsar with an ultramassive ($M$ $>$1.14$M_{solar}$) WD companion~\citep{Edwards2001}. The polarization profile for PSR J1157$-$5112 shows a single peak at 1369 MHz in Fig.4. The profile is linearly polarized, and there is little or no variation in the PA. No significant circular polarization is observed across the whole profile.

\subsubsection{\textbf{PSR J1420$-$5416(PSR B1417$-$54)}}
PSR J1420$-$5416 was first discovered at 408 MHz in Molonglo survey \citep{Manchester1978}. Fig.3 shows the pulse profile of the pulsar consists of two closely spaced components at 2368 MHz. The leading component is less linear polarized than the trailing component, and the PA is a continuous decrease across the trailing component. There is significant right-hand circular polarization.

\subsubsection{\textbf{PSR J1615$-$2940(PSR B1612$-$29)}}
Similar to PSR J1420$-$5416, PSR J1615$-$2940 was also discovered in the second Molonglo pulsar survey \citep{Manchester1978}. Fig.3 shows a pulse profile with two components at 2368 MHz. The pulse profile is very low in linear polarization, and thus there are very few PA measured. There is a sense reversal of the circular polarization from right-hand to left-hand under the trailing part of the profiles.

\subsubsection{\textbf{PSR J1728$-$0007(PSR B1726$-$00)}}
PSR J1728$-$0007 was found in a survey for short period pulsars~\citep{Stokes1985}. \citet{Weisberg1999} published its profile at 1418 MHz, but the linear polarization is too weak to measure. Fig.3 shows the pulse profile consists of two components at 2368 MHz. The pulsar is linearly polarized with a continuous increasing PA across the profile.

\subsubsection{\textbf{PSR J1746$-$2856}}
PSR J1746$-$2856, a highly dispersed pulsar, was discovered in the direction of the Galactic Centre at 3.1 GHz with the Parkes radio telescope \citep{Johnston2006}. The mean pulse profiles and polarization parameters for this pulsar at 2368 MHz are shown in Fig.3. The degree of linear polarization is low, with very few PA measured. There are probably four or five distinct components, but it is impossible to classify this pulsar without additional information.

\subsubsection{\textbf{PSR J1805$+$0306(PSR B1802$+$03)}}
Polarization profiles at 2368 MHz and 1369 MHz for PSR J1805+0306 are presented in Figure 2. The profiles show two components at both frequencies. The leading part of the profile has relatively high fractional linear polarization, whereas the trailing part of the profile is essentially unpolarized at both frequencies. There is a shallow rotation of the PA through the leading component at 2368 MHz. The polarization profiles at 1369 MHz are in good agreement with those presented by  ~\citet{Weisberg1999} at Arecibo 1418 MHz.

\subsubsection{\textbf{PSR J1810$-$5338(PSR B1806$-$53)}}
As shown in Fig.2, the mean pulse profiles of PSR J1810-5338 are almost the same at the two frequencies. There are three pulse components of the pulsar, which are similar to that at 660 MHz ~\citep{Jiang2014}. The leading and trailing components are weaker than the central component of the pulse profile. The central component is less linear polarized than the leading and trailing component at both frequencies. Flat increasing PA swings through the overall profiles can be seen both at 2368 and 1369 MHz. There is a sense reversal of the circular polarization from right-hand to left-hand under the whole profiles at both frequencies.

\subsubsection{\textbf{PSR J1833$-$6023(PSR B1828$-$60)}}
The mean pulse profiles and polarization parameters for PSR J1833$-$6023 at 2368 and 1369 MHz are similar, which are shown in Fig.2. The pulsar was first discovered at a frequency of 408 MHz by the second Molonglo pulsar survey ~\citep{Manchester1978}. PSR J1833$-$6023 has a profile with only one narrow component, and the profile is linearly polarized. There is a strong sense reversal of the circular polarization from right-hand to left-hand under the overall pulse, and a steep decreasing PA swing across it.

\subsubsection{\textbf{PSR J1848$-$1952(PSR B1845$-$19)}}
PSR J1848$-$1952 was first discovered by an extensive survey for pulsars which had been undertaken using observations at the Molonglo Radio Observatory and the Australian National Radio Astronomy Observatory, Parkes. The observing frequency for both the Molonglo and Parkes observations was 408 MHz ~\citep{Manchester1978}. As shown in Fig.2, the mean pulse profile of PSR J1848$-$1952 at both 2368 MHz and 1369 MHz consists of two close sharp components. The PA variations are different, but the linear and circular polarizations are almost the same at two frequencies. The overall pulse has obvious linear polarization, and the trailing component is less linear polarized than the leading component. A high degree of right-hand circular polarization was observed across the whole profile. There is a continuous decreasing PA across the profile at 2368 MHz, while the PA variation is complex with a shallow rotation swing through it at 1369 MHz.

\subsubsection{\textbf{PSR J1854$-$1421(PSR B1851$-$14)}}
Similar to PSR J1848$-$1952, PSR J1854$-$1421 was first also discovered by the second Molonglo pulsar survey ~\citep{Manchester1978}. The pulse profile of the pulsar in Fig.2 shows a single peak at both 2368 MHz and 1369 MHz and is linearly polarized. There is a continuous decreasing PA across the profile at 1369 MHz, while the PA variation is discontinuous decreasing with a PA jump at 2368 MHz. No significant circular polarization is observed across the whole profile.

\subsubsection{\textbf{PSR J1903$-$0632(PSR B1900$-$06)}}
Mean pulse profiles and polarization parameters for PSR J1903$-$0632 at 2368 MHz and 1369 MHz are very similar, shown in Fig.3. The pulsar was discovered during a systematic search for pulsars at a frequency of 408 MHz, carried out with the Jodrell Bank Mark IA radio telescope in 1972 ~\citep{Davies1972}. Similar to 408 MHz, the pulse profile of the pulsar also has two pulse components at two frequencies. The trailing component is less linear polarized than the leading component. PSR J1903$-$0632 shows a small amount of circular polarization from left-hand to right-hand under the overall profile. The PA variation across two components appears continuous and has a negative slope.

\subsubsection{\textbf{PSR J2222$-$0137}}
PSR J2222$-$0137, a $2.44$ day binary pulsar with a massive CO WD companion, was first discovered in the Green Bank Telescope (GBT) 350 MHz drift scan survey. The mean pulse profile and polarization parameters for PSR J2222$-$0137 at 2368 MHz given in Fig.3 have far more details than previously published results at 820 MHz ~\citep{Boyles2013}. The pulse profile is very similar at both frequencies, with only one pulse component. The linear polarization is low under the overall profile. There is a hint of circular polarization from left-hand to right-hand against the profile. The PA variation across the profile is complex with regions of increasing and decreasing PAs, and a PA jump can be seen at phase $0.504$.

\subsection{Correlations of spectral index with pulsar parameters}\label{Sect.3.4}

We test for a correlation between  spectral index $\alpha$ and $\lg|x|$ for all the 18 pulsars, where $x$ is one of the pulsar parameters below. The parameters are spin frequency $\nu$, spin-down rate $\dot{\nu}$, the magnetic field at the light cylinder radius $B_{LC}$, the characteristic age $\tau_{c}$, the surface magnetic field $B_{surf}$, the spin-down luminosity $\dot{E}$, pulse period $P$, and period derivative $\dot{P}$ of the pulsar. We took all values of pulsars parameters from the ATNF pulsar catalog. Most are covariant because these quantities depend on basic pulsar parameters, such as pulse period and its derivative.

We first measured the correlation by visual inspection, then computed the Spearman rank correlation coefficient to characterize its strength. We test all pulsars in our single power-law data set first, then the isolated pulsars. There are only two pulsars in the binary system. The correlation coefficients, corresponding $p-$values, and the number of pulsars N for which the computed correlation are shown in Table 3.

We find a mildly correlation between the spectral index and the pulse period for normal pulsars. All the other combinations show no correlation. For isolated pulsars, we found a mildly negative correlation of spectral index with the surface magnetic field.

\section{Discussion and Conclusions}\label{Sect.4}

The minimum detectable flux density ($S_{min}$) of the UWL system can be evaluated using the radiometer equation ~\citep{Manchester1996}:
\begin{equation}
S_{min}=\frac{\alpha \beta T_{sys}}{G(N_{p}\Delta\nu T)^{1/2}}(\frac{W}{P-W})^{1/2}
\end{equation}
where $\alpha=S/N$, $\beta$ is the factor of digitization and other processing losses, $T_{sys}$ is the sum of receiver temperature and sky temperature ($K$), $G$ is the gain of the telescope, $N_{p}$ is the number of polarization (two in this case), $\Delta\nu$ is the observing bandwidth (MHz), $T$ is the observation length ($s$), $W$ is the effective pulse width in time units, and $P$ is the pulse period ~\citep{Dewey1985}. According to Eq.1, the minimum detectable flux density of the UWL system varies from pulsar to pulsar with the different of $W$, $P$, and $T$. Here, we adopt $\alpha=10$, $\beta=1.5$~\citep{Manchester1996}, $G=0.64K Jy^{-1}$~\citep{Edwards2001}, $T_{sys}=22K$ and $\Delta\nu=3328$ MHz ~\citep{Hobbs2020}. The minimum detectable flux density ($S_{min}$) of the UWL system ranges from $0.02$ to $0.76$ $mJy$ for pulsars with different duty cycle.

However, 54 pulsars have been observed and only 25 have been detected. Among the 29 pulsars undetected, six pulsars are MSPs with periods from $1.7$ to $3.7$ $ms$, and the periods of the other 23 pulsars range from $44ms$ to $4.3s$. We estimated the flux densities at 2368 MHz assuming the spectral index $\alpha = -1.6 $ for undetected pulsars with only one published flux density (the flux densities of PSR J1231-1411 and J2256-1024 were estimated separately). As shown in Table B.1, all the estimated S$_{2368}$ of 22 pulsars are potentially detectable by the UWL system.

\textbf{—}  Eq.1 is not suitable for estimating the sensitivity of the UWL system. Because the bandwidth of the UWL system is very wide, the flux densities of pulsars in the bandwidth change significantly and decrease in the high-frequency part.

\textbf{—}  The flux density of a radio pulsar decreases rapidly with increasing observing frequency. According to previous statistics, the spectral index of pulsars varies significantly. As shown in Table B.1, the estimated flux density at 2368 MHz for some pulsars is significantly higher than the detection threshold of the UWL system. Those undetected pulsars may have a steeper radio spectrum.

\textbf{—} The bandwidth of the UWL system can not be fully utilized. Part of the bandwidth is contaminated by RFI and has to be removed. Therefore, the sensitivity of the UWL system is not as high as we calculated using Eq.1. In fact, most of the frequency channels from 700 MHz to 1000 MHz have been removed, and the radio emission of pulsars is bright at the low end of the UWL system.

\textbf{—} The radio emission of pulsars is strongly affected by propagation through the interstellar medium (ISM) ~\citep{Kumamoto2021}. The typical observing length of our observations is only 10 minutes. But the timescale of diffractive scintillation could be hours which is much longer than our observing length.

The correlation coefficients for pulsars in our sample, along with previous results from \citet{Zhao2019}, \citet{Jankowski2017}, \citet{Han2016}, and \citet{Lorimer1995} are listed in Table 3. \citet{Jankowski2017} obtained Spearman rank correlations of the spectral index with various pulsar
parameters for 276 pulsars with simple power-law spectra. They found moderate correlations with $\dot{\nu}$, $B_{LC}$ and $\dot{E}$ for young pulsars. \citet{Han2016} and \citet{Lorimer1995} found a very weak correlation of the spectral index with $\dot{E}$. For pulsars in our sample, we find moderate correlations between spectral index with $\nu$, which are in the sense of steeper spectra for fast-rotating pulsars. No consistent correlation is found across the different samples, and results from disparate samples are distinct.

In this work, we have carried out ultra-wide bandwidth observations of 19 pulsars with the Parkes telescope. Flux density measurements, spectral properties, polarization profiles, and pulse widths at 2368 MHz or 1369 MHz have been presented. The non-detection polarization of 12 out of 25 pulsars maybe not be surprising. As shown in Fig.A1, the S/N ratio of some pulse profiles (PSRs J0057$-$7201, J1721$-$1939, J1816$-$5548, etc.) is low. It is not very likely to measure polarization with a low S/N ratio. Other pulse profiles have a medium S/N ratio. The degree of linear polarization for these pulsars may be low. To measure the pulse profile, polarization and RM with high precision and study their evolution with frequency, longer observations for these pulsars are needed. We also noted that there are a large number of pulsars in the catalog that still do not have such basic information as the flux density near 1.4 GHz. Observing each pulsar for a longer time or using telescopes with larger observation apertures (such as FAST) at a specific frequency is needed when we continue this work in the future.

\begin{acknowledgements}
The Parkes radio telescope is part of the ATNF which is funded by the Commonwealth of Australia for operation as a National Facility managed by CSIRO. This paper includes archived data obtained through the CSIRO Data Access Portal (http://data.csiro.au). This work is supported by the National Natural Science Foundation of China (No.NSFC12041304), National SKA Program of China (No. 2020SKA0120100), Youth Innovation Promotion Association of Chinese Academy of Sciences, National Key Research and Development Program of China (No.2017YFA0402602), the CAS Jianzhihua project, and Heaven Lake Hundred-Talent Program of Xinjiang Uygur Autonomous Region of China.

\end{acknowledgements}

\begin{appendix}

\section{Figures of averaged pulse profiles}
Averaged pulse profiles at 2368 MHz or 1369 MHz were obtained with Parkes telescope for 16 pulsars, five of them were obtained at both frequencies.

\setcounter{figure}{0}
\begin{figure}
\centering
\includegraphics[width=6cm, height=7cm, angle=-90]{J0418.jpg}
\includegraphics[width=6cm, height=7cm, angle=-90]{J1057.jpg}
\includegraphics[width=6cm, height=7cm, angle=-90]{J1423.jpg}
\includegraphics[width=6cm, height=7cm, angle=-90]{2J1527.jpg}
\includegraphics[width=6cm, height=7cm, angle=-90]{2J1625.jpg}
\includegraphics[width=6cm, height=7cm, angle=-90]{J1721.jpg}
\includegraphics[width=6cm, height=7cm, angle=-90]{J1816.jpg}
\caption{Averaged pulse profiles at 2368 MHz were obtained with Parkes telescope for 7 pulsars.}
\label{f1}
\end{figure}

\begin{figure}
\centering
\includegraphics[width=6cm, height=7cm, angle=-90]{dJ0057.jpg}
\includegraphics[width=6cm, height=7cm, angle=-90]{dJ0418.jpg}
\includegraphics[width=6cm, height=7cm, angle=-90]{dJ0458.jpg}
\includegraphics[width=6cm, height=7cm, angle=-90]{dJ1057.jpg}
\includegraphics[width=6cm, height=7cm, angle=-90]{dJ1423.jpg}
\includegraphics[width=6cm, height=7cm, angle=-90]{2dJ1510.jpg}
\includegraphics[width=6cm, height=7cm, angle=-90]{2dJ1527.jpg}
\includegraphics[width=6cm, height=7cm, angle=-90]{dJ1539.jpg}
\label{f2}
\end{figure}

\begin{figure}
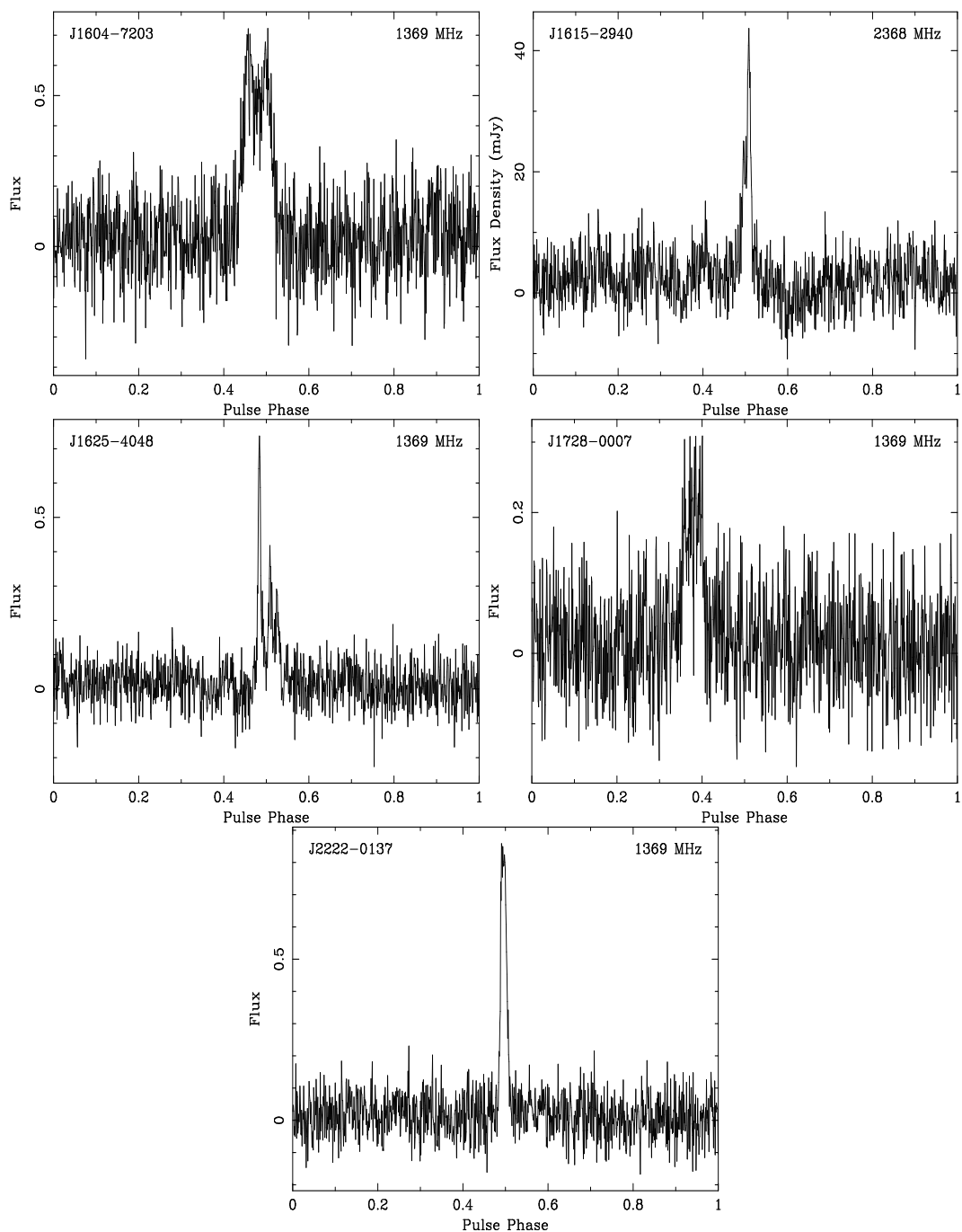

\centering
\includegraphics[width=6cm, height=7cm, angle=-90]{dJ1604.jpg}
\includegraphics[width=6cm, height=7cm, angle=-90]{2J1615.jpg}
\includegraphics[width=6cm, height=7cm, angle=-90]{dJ1625.jpg}
\includegraphics[width=6cm, height=7cm, angle=-90]{2dJ1728.jpg}
\includegraphics[width=6cm, height=7cm, angle=-90]{dJ2222.jpg}
\caption{Averaged pulse profiles of 13 pulsars at a center frequency of 1369 MHz were obtained with Parkes telescope.}
\label{f2}
\end{figure}

\section{Estimated flux densities at 2368 MHz ($S_{2368}$) of 22 Pulsars and pulse widths at other frequencies for 17 pulsars}
Estimated flux densities of 22 Pulsars were estimated at 2368 MHz ($S_{2368}$). Pulse widths for 17 Pulsars were measured previously at other frequencies.

\begin{table}
\begin{center}
\caption[]{Estimated flux densities at 2368 MHz ($S_{2368}$) for 22 Pulsars\label{a1}}
\setlength{\tabcolsep}{4pt}
\small
 \begin{tabular}{ccccccccccl}
  \hline\noalign{\smallskip}
PSR J & $S_{150}$& $S_{300}$&$S_{400}$ & $S_{600}$ & $S_{800}$ &$S_{1400}$ & $S_{2000}$ &$\alpha$ &$S_{2368}$& References\\
$\qquad$ &  (mJy) &  (mJy) & (mJy) &  (mJy) & (mJy) & (mJy)& (mJy)& &(mJy)&for position\\
\hline
J0458$-$0505& & & & & 0.6& & & $-1.6$ &0.11 & \citet{Lynch2013}\\
J0502$-$6617& & & 1.0 & & & & & $-1.6$ & 0.06&\citet{Lorimer1995}\\
J0600$-$5756& & &2.1 & & & & & $-1.6$ & 0.12&\citet{Taylor1993}\\
J0614$-$3329& & & & & 1.5 & & & $-1.6$ & 0.26&\citet{Ray2011}\\
J0702$-$4956& &15.7 & & & & & & $-1.6$ & 0.58&\citet{Bhattacharyya2016}\\
J0946$+$0951& & &4(1)& & & & & $-1.6$ & 0.23&\citet{Lorimer1995}\\
J1156$-$5909& & &7 & & & & & $-1.6$ &0.41&\citet{Lyne1998}\\
J1227$-$4853& & & &6.6(2)& & & & $-1.6$ &0.73&\citet{Roy2015}\\
J1231$-$1411&2.2 & & & &0.4& & & $-1.02$& 0.13&\citet{Ransom2011}\\
J1232$-$4742& & & & & & 2.38(6)& & $-1.6$ &1.03 &\citet{Xie2019}\\
J1402$-$5124& & & 10 & & & & & $-1.6$ &0.58& \citet{Manchester1978}\\
J1510$-$4422& & & 14 & & & & & $-1.6$ &0.81 &\citet{Taylor1993}\\
J1604$-$7203& & & 10 & & & & & $-1.6$ &0.58 &\citet{Lyne1998}\\
J1704$-$6016& & & 23 & & & & & $-1.6$ &1.34&\citet{Taylor1993}\\
J1745$-$2910& & & & & & &1.2 & $-1.6$ &0.92& \citet{Deneva2009}\\
J1745$-$2912& & & & & & &0.2 & $-1.6$ &0.15 &\citet{Johnston2006}\\
J1809$-$3547& & & 21 & & & & & $-1.6$ & 1.22&\citet{Lyne1998}\\
J1947$-$4215& & & 7 & & & & & $-1.6$ & 0.41&\citet{Lyne1998}\\
J2012$-$2029& & & & & 1.00(12)& & & $-1.6$ & 0.18&\citet{Boyles2013}\\
J2013$-$0649& & & & & 0.8 & & & $-1.6$ &0.15 &\citet{Lynch2013}\\
J2033$-$1938& & & & & 1.13 & & & $-1.6$ & 0.20&\citet{Boyles2013}\\
J2256$-$1024& & & 13 & & & 0.73 & & $-2.30$ & 0.22&\citet{Crowter2020}\\
  \noalign{\smallskip}\hline
\end{tabular}
\end{center}
\end{table}

\begin{table}
\begin{center}
\caption[]{Pulse widths from literature for 17 Pulsars\label{a2}}
\setlength{\tabcolsep}{2.5pt}
\small
 \begin{tabular}{ccccl}
  \hline\noalign{\smallskip}
PSR J &Frequency& $W_{50}$&$W_{10}$ &References\\
$\qquad$&(MHz)&(deg)&(deg)&for position\\
\hline
J0348$+$0432&350&27 &50.4&~\citet{McEwen2020}\\
J0418$-$4154&843&7.1 &13.2&~\citet{Jankowski2019}\\
J1057$-$4754&1374&11.4 &&~\citet{Edwards2001}\\
J1420$-$5416&400&9.6 &16.1&~\citet{Taylor1993}\\
J1423$-$6953&1374&3.6 &&~\citet{Edwards2001}\\
J1527$-$3931&350&6.0 &9.7&~\citet{McEwen2020}\\
J1615$-$2940&408&3.7 &9.2&~\citet{Lorimer1995}\\
J1625$-$4048&400&15(1) &25(2)&~\citet{D'Amico1998}\\
J1728$-$0007&408&13.3 &25.3&~\citet{Lorimer1995}\\
J1805$+$0306&350&7 &23&~\citet{McEwen2020}\\
J1810$-$5338&660&10 &30&~\citet{Qiao1995}\\
J1816$-$5643&1374&10.8 &37.0&~\citet{Jacoby2009}\\
J1833$-$6023&400&11 &15&~\citet{Taylor1993}\\
J1848$-$1952&408&22.1 &31.7&~\citet{Lorimer1995}\\
J1854$-$1421&408&5.9 &14.7&~\citet{Lorimer1995}\\
J1903$-$0632&408&13.4 &31.4&~\citet{Lorimer1995}\\
J2222$-$0137&350&12 &24&~\citet{McEwen2020}\\
  \noalign{\smallskip}\hline
\end{tabular}
\end{center}
\end{table}

\end{appendix}

\label{lastpage}

\clearpage


\begin{thebibliography}{18}
\providecommand{\natexlab}[1]{#1}
\providecommand{\selectlanguage}[1]{\relax}

\bibitem[{{An} et~al.(2017){An}, {Chen}, {Mohan}, \& {Lao}}]{An2017}
{An}, T., {Chen}, X., {Mohan}, P., \& {Lao}, B.~Q. 2017, Acta Astronomica
  Sinica, 58, 43

\bibitem[{{Baan}(2011)}]{Baan2011}
{Baan}, W.~A. 2011, in 2011 XXXth URSI General Assembly and Scientific
  Symposium, 1--2

\bibitem[{Baek et~al.(2015)Baek, Park, Ahn, \& Choo}]{Baek2015}
Baek, S.-J., Park, A., Ahn, Y.-J., \& Choo, J. 2015, The Analyst, 140 1, 250

\bibitem[{{Boonstra}(2005)}]{Boonstra2005}
{Boonstra}, A.-J. 2005, {Radio frequency interference mitigation in radio
  astronomy}, Ph.D. thesis, -

\bibitem[{Ford \& Buch(2014)}]{Ford2014}
Ford, J.~M., \& Buch, K.~D. 2014, in 2014 IEEE Geoscience and Remote Sensing
  Symposium, 231--234

\bibitem[{{Fridman} \& {Baan}(2001)}]{Fridman2001}
{Fridman}, P.~A., \& {Baan}, W.~A. 2001, \aap, 378, 327

\bibitem[{{Han} et~al.(2021){Han}, {Wang}, {Wang} et~al.}]{Han2021}
{Han}, J.~L., {Wang}, C., {Wang}, P.~F., et~al. 2021, Research in Astronomy and
  Astrophysics, 21, 107

\bibitem[{Hu et~al.(2021)Hu, Li, Wang et~al.}]{hu2021}
Hu, W., Li, Y., Wang, Y., et~al. 2021, \emph{1/f Noise Analysis for FAST HI
  Intensity Mapping Drift-Scan Experiment}

\bibitem[{{Jiang} et~al.(2020){Jiang}, {Tang}, {Hou} et~al.}]{Jiang2020}
{Jiang}, P., {Tang}, N.-Y., {Hou}, L.-G., et~al. 2020, Research in Astronomy
  and Astrophysics, 20, 064

\bibitem[{{Jiang} et~al.(2019){Jiang}, {Yue}, {Gan} et~al.}]{Jiang2019}
{Jiang}, P., {Yue}, Y., {Gan}, H., et~al. 2019, Science China Physics,
  Mechanics, and Astronomy, 62, 959502

\bibitem[{{Kesteven}(2005)}]{Kesteven2005}
{Kesteven}, M. 2005, in Proceedings. (ICASSP '05). IEEE International
  Conference on Acoustics, Speech, and Signal Processing, 2005., vol.~5,
  v/873--v/876 Vol. 5

\bibitem[{{Nan} et~al.(2011){Nan}, {Li}, {Jin} et~al.}]{Nan2011}
{Nan}, R., {Li}, D., {Jin}, C., et~al. 2011, International Journal of Modern
  Physics D, 20, 989

\bibitem[{{Qian} et~al.(2020){Qian}, {Yao}, {Sun} et~al.}]{Qian2020}
{Qian}, L., {Yao}, R., {Sun}, J., et~al. 2020, The Innovation, 1, 100053

\bibitem[{{Wang} et~al.(2021){Wang}, {Zhang}, {Hu} et~al.}]{Wang2021}
{Wang}, Y., {Zhang}, H.-Y., {Hu}, H., et~al. 2021, Research in Astronomy and
  Astrophysics, 21, 018

\bibitem[{{Zeng} et~al.(2021){Zeng}, {Chen}, {Li} et~al.}]{Zeng2021}
{Zeng}, Q., {Chen}, X., {Li}, X., et~al. 2021, \mnras, 500, 2969

\bibitem[{{Zhang} et~al.(2021{\natexlab{a}}){Zhang}, {Xu}, {Li}
  et~al.}]{Zhang2021}
{Zhang}, C.-P., {Xu}, J.-L., {Li}, G.-X., et~al. 2021{\natexlab{a}}, Research
  in Astronomy and Astrophysics, 21, 209

\bibitem[{{Zhang} et~al.(2020){Zhang}, {Wu}, {Yue} et~al.}]{Zhang2020}
{Zhang}, H.-Y., {Wu}, M.-C., {Yue}, Y.-L., et~al. 2020, Research in Astronomy
  and Astrophysics, 20, 075

\bibitem[{{Zhang} et~al.(2021{\natexlab{b}}){Zhang}, {Ren}, {Li}, {Nguyen}, \&
  {Stoica}}]{Zhang2021rfi}
{Zhang}, T., {Ren}, J., {Li}, J., {Nguyen}, L.~H., \& {Stoica}, P.
  2021{\natexlab{b}}, arXiv e-prints, arXiv:2102.08987

\end{thebibliography}


\begin{thebibliography}{99}

\bibitem[Hewish et al.(1968)]{Hewish1968} Hewish, A., Bell, S.~J., Pilkington, J.~D.~H., et al.\ 1968, \nat, 217, 709

\bibitem[Jankowski et al.(2018)]{Jankowski2017} Jankowski, F., van Straten, W., Keane, E.~F., et al.\ 2018, \mnras, 473, 4436

\bibitem[Goldreich \& Julian(1969)]{Gold1969} Goldreich, P. \& Julian, W.~H.\ 1969, \apj, 157, 869

\bibitem[Backer(1972)]{Backer1972} Backer, D.~C.\ 1972, \apjl, 174, L157

\bibitem[Backer(1976)]{Backer1976} Backer, D.~C.\ 1976, \apj, 209, 895

\bibitem[Bhattacharyya et al.(2016)]{Bhattacharyya2016} Bhattacharyya, B., Cooper, S., Malenta, M., et al.\ 2016, \apj, 817, 130

\bibitem[Boyles et al.(2013)]{Boyles2013} Boyles, J., Lynch, R.~S., Ransom, S.~M., et al.\ 2013, \apj, 763, 80

\bibitem[Chen \& Wang(2014)]{Chen2014} Chen, J.~L. \& Wang, H.~G.\ 2014, \apjs, 215, 11

\bibitem[Crowter et al.(2020)]{Crowter2020} Crowter, K., Stairs, I.~H., McPhee, C.~A., et al.\ 2020, \mnras, 495, 3052

\bibitem[Dai et al.(2015)]{Dai2015} Dai, S., Hobbs, G., Manchester, R.~N., et al.\ 2015, \mnras, 449, 3223

\bibitem[D'Amico et al.(1998)]{D'Amico1998} D'Amico, N., Stappers, B.~W., Bailes, M., et al.\ 1998, \mnras, 297, 28

 \bibitem[Davies et al.(1972)]{Davies1972} Davies, J.~G., Lyne, A.~G., \& Seiradakis, J.~H.\ 1972, \nat, 240, 229

 \bibitem[Deneva et al.(2009)]{Deneva2009} Deneva, J.~S., Cordes, J.~M., \& Lazio, T.~J.~W.\ 2009, \apjl, 702, L177

\bibitem[Dewey et al.(1985)]{Dewey1985} Dewey, R.~J., Taylor, J.~H., Weisberg, J.~M., et al.\ 1985, \apjl, 294, L25

\bibitem[Edwards et al.(2001)]{Edwards2001} Edwards, R.~T., Bailes, M., van Straten, W., et al.\ 2001, \mnras, 326, 358

\bibitem[Gould \& Lyne(1998)]{Gould1998} Gould, D.~M. \& Lyne, A.~G.\ 1998, \mnras, 301, 235

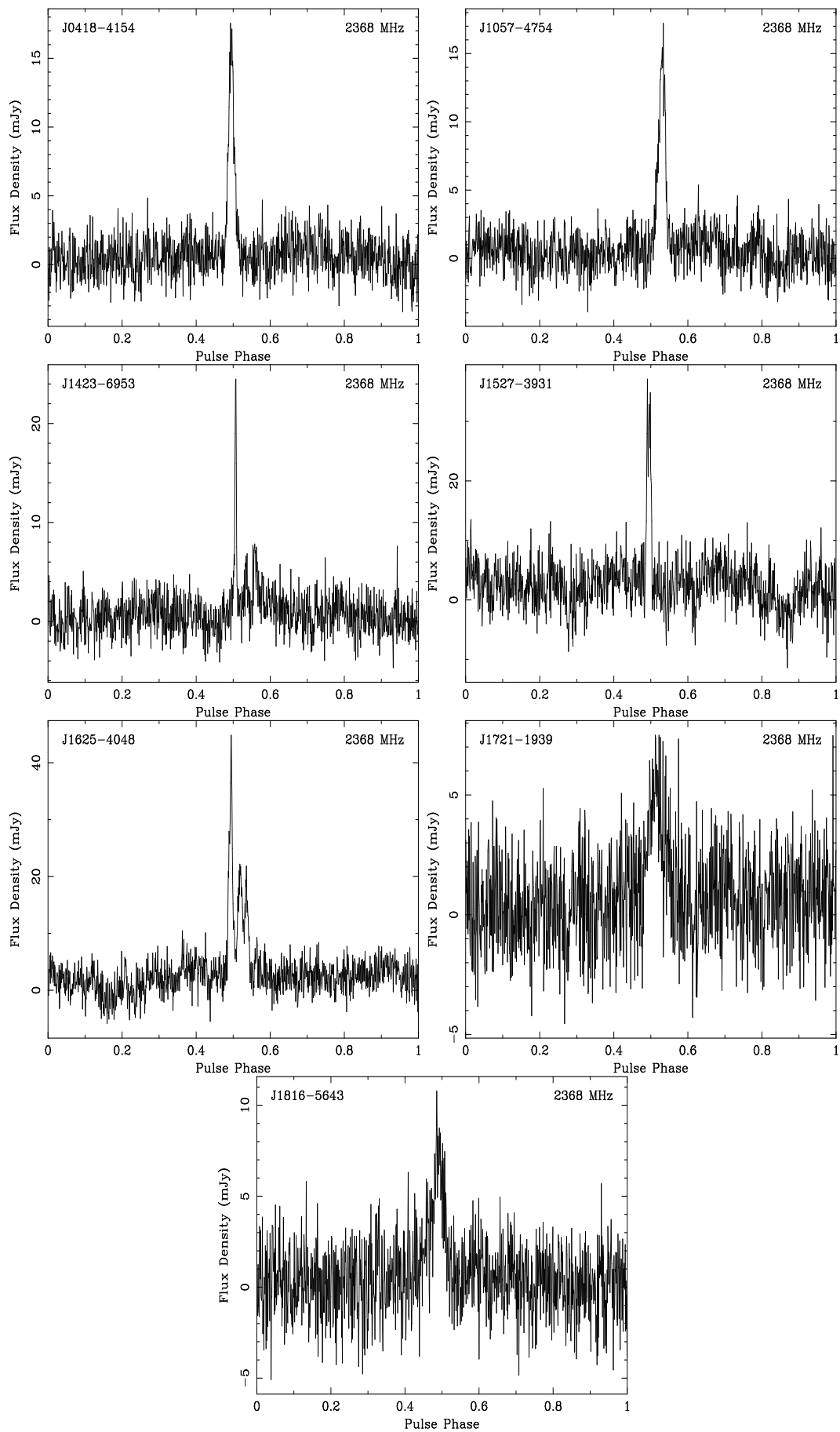
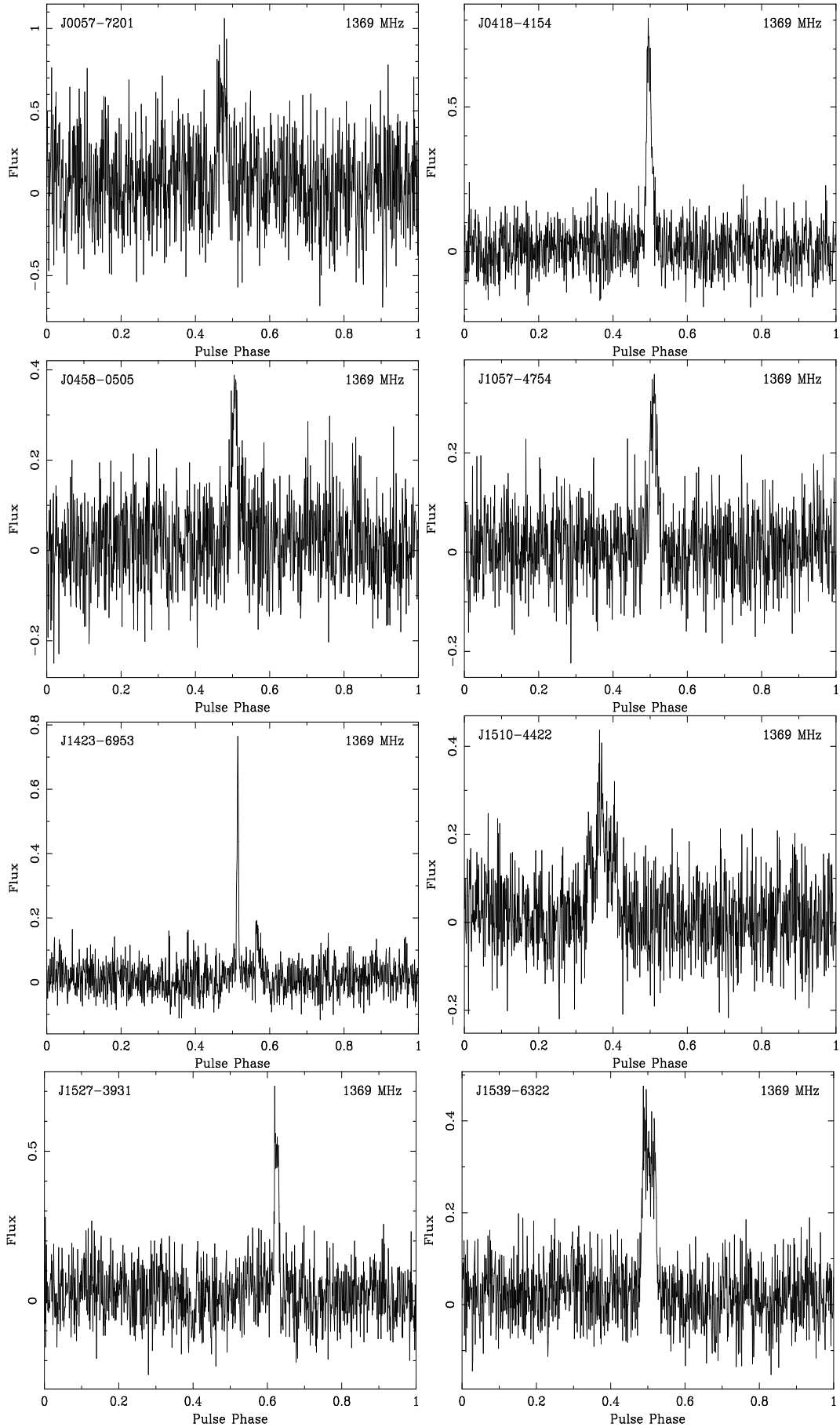
\bibitem[Han et al.(2016)]{Han2016} Han, J., Wang, C., Xu, J., et al.\ 2016, Research in Astronomy and Astrophysics, 16, 159

\bibitem[Hobbs et al.(2011)]{Hobbs2011} Hobbs, G., Miller, D., Manchester, R.~N., et al.\ 2011, PASA, 28, 202

\bibitem[Hobbs et al.(2020)]{Hobbs2020} Hobbs, G., Manchester, R.~N., Dunning, A., et al.\ 2020, PASA, 37, e012

\bibitem[Hotan et al.(2014)]{Hotan2014} Hotan, A.~W., Bunton, J.~D., Harvey-Smith, L., et al.\ 2014, PASA, 31, e041

\bibitem[Izvekova et al.(1981)]{Izvekova1981} Izvekova, V.~A., Kuzmin, A.~D., Malofeev, V.~M., et al.\ 1981, \apss, 78, 45

\bibitem[Jacoby et al.(2009)]{Jacoby2009} Jacoby, B.~A., Bailes, M., Ord, S.~M., et al.\ 2009, \apj, 699, 2009

 \bibitem[Jankowski et al.(2019)]{Jankowski2019} Jankowski, F., Bailes, M., van Straten, W., et al.\ 2019, \mnras, 484, 3691

\bibitem[Jiang \& You(2014)]{Jiang2014} Jiang, X. \& You, X.-P.\ 2014, CAA, 38, 32

\bibitem[Johnston et al.(2006)]{Johnston2006} Johnston, S., Kramer, M., Lorimer, D.~R., et al.\ 2006, \mnras, 373, L6

\bibitem[Kumamoto et al.(2021)]{Kumamoto2021} Kumamoto, H., Dai, S., Johnston, S., et al.\ 2021, \mnras, 501, 4490

\bibitem[Levin et al.(2013)]{Levin2013} Levin, L., Bailes, M., Barsdell, B.~R., et al.\ 2013, \mnras, 434, 1387

\bibitem[Lommen et al.(2000)]{Lommen2000} Lommen, A.~N., Zepka, A., Backer, D.~C., et al.\ 2000, \apj, 545, 1007

\bibitem[Lorimer et al.(1995)]{Lorimer1995} Lorimer, D.~R., Yates, J.~A., Lyne, A.~G., et al.\ 1995, \mnras, 273, 411

\bibitem[Lynch et al.(2013)]{Lynch2013} Lynch, R.~S., Boyles, J., Ransom, S.~M., et al.\ 2013, \apj, 763, 81

\bibitem[Lyne \& Smith(1968)]{Lyne1968} Lyne, A.~G. \& Smith, F.~G.\ 1968, \nat, 218, 124. doi:10.1038/218124a0

\bibitem[Lyne et al.(1998)]{Lyne1998} Lyne, A.~G., Manchester, R.~N., Lorimer, D.~R., et al.\ 1998, \mnras, 295, 743

\bibitem[Malofeev \& Malov(1980)]{Malofeev1980} Malofeev, V.~M. \& Malov, I.~F.\ 1980, \sovast, 24, 54

\bibitem[Malofeev et al.(2000)]{Malofeev2000} Malofeev, V.~M., Malov, O.~I., \& Shchegoleva, N.~V.\ 2000, Astronomy Reports, 44, 436

\bibitem[Manchester et al.(2005)]{Manchester2005} Manchester, R.~N., Hobbs, G.~B., Teoh, A., et al.\ 2005, VizieR Online Data Catalog, VII/245

\bibitem[Manchester et al.(1978)]{Manchester1978} Manchester, R.~N., Lyne, A.~G., Taylor, J.~H., et al.\ 1978, \mnras, 185, 409

\bibitem[Manchester et al.(1996)]{Manchester1996} Manchester, R.~N., Lyne, A.~G., D'Amico, N., et al.\ 1996, \mnras, 279, 1235

\bibitem[Manchester et al.(2013)]{Manchester2013} Manchester, R.~N., Hobbs, G., Bailes, M., et al.\ 2013, PASA, 30, e017

\bibitem[Maron et al.(2000)]{Maron2000} Maron, O., Kijak, J., Kramer, M., et al.\ 2000, \aaps, 147, 195

\bibitem[McEwen et al.(2020)]{McEwen2020} McEwen, A.~E., Spiewak, R., Swiggum, J.~K., et al.\ 2020, \apj, 892, 76

\bibitem[Phillips \& Wolszczan(1992)]{Phillips1992} Phillips, J.~A. \& Wolszczan, A.\ 1992, \apj, 385, 273

\bibitem[Qiao et al.(1995)]{Qiao1995} Qiao, G., Manchester, R.~N., Lyne, A.~G., et al.\ 1995, \mnras, 274, 572

\bibitem[Radhakrishnan \& Cooke(1969)]{Radhakrishnan1969} Radhakrishnan, V. \& Cooke, D.~J.\ 1969, \aplett, 3, 225

\bibitem[Rankin(1983)]{Rankin1983} Rankin, J.~M.\ 1983, \apj, 274, 333

\bibitem[Rankin et al.(1970)]{Rankin1970} Rankin, J.~M., Comella, J.~M., Craft, H.~D., et al.\ 1970, \apj, 162, 707

\bibitem[Ransom et al.(2011)]{Ransom2011} Ransom, S.~M., Ray, P.~S., Camilo, F., et al.\ 2011, \apjl, 727, L16

\bibitem[Ray et al.(2011)]{Ray2011} Ray, P.~S., Kerr, M., Parent, D., et al.\ 2011, \apjs, 194, 17

\bibitem[Roy et al.(2015)]{Roy2015} Roy, J., Ray, P.~S., Bhattacharyya, B., et al.\ 2015, \apjl, 800, L12

\bibitem[Ruderman \& Sutherland(1975)]{Ruderman1975} Ruderman, M.~A. \& Sutherland, P.~G.\ 1975, \apj, 196, 51

\bibitem[Sieber(1973)]{Sieber1973} Sieber, W.\ 1973, \aap, 28, 237

\bibitem[Stokes et al.(1985)]{Stokes1985} Stokes, G.~H., Taylor, J.~H., Welsberg, J.~M., et al.\ 1985, \nat, 317, 787

\bibitem[Taylor et al.(1993)]{Taylor1993} Taylor, J.~H., Manchester, R.~N., \& Lyne, A.~G.\ 1993, \apjs, 88, 529

\bibitem[Thorsett(1991)]{Thorsett1991} Thorsett, S.~E.\ 1991, \apj, 377, 263

\bibitem[Weisberg et al.(1999)]{Weisberg1999} Weisberg, J.~M., Cordes, J.~M., Lundgren, S.~C., et al.\ 1999, \apjs, 121, 171

\bibitem[Xie et al.(2019)]{Xie2019} Xie, Y.-W., Wang, J.-B., Hobbs, G., et al.\ 2019, Research in Astronomy and Astrophysics, 19, 103

\bibitem[Xilouris et al.(1996)]{Xilouris1996} Xilouris, K.~M., Kramer, M., Jessner, A., et al.\ 1996, \aap, 309, 481

\bibitem[Yan et al.(2011)]{Yan2011} Yan, W.~M., Manchester, R.~N., van Straten, W., et al.\ 2011, \mnras, 414, 2087

\bibitem[Zhang et al.(2019)]{Zhang2019} Zhang, L., Hobbs, G., Manchester, R.~N., et al.\ 2019, \apjl, 885, L37

\bibitem[Zhao et al.(2019)]{Zhao2019} Zhao, R.-S., Yan, Z., Wu, X.-J., et al.\ 2019, \apj, 874, 64

\end{thebibliography}
\end{document}